\documentclass[11pt]{revtex4}

\usepackage{multirow}
\usepackage{subfigure, amsfonts, amsmath}
\usepackage{hyperref}
\usepackage{longtable}
\usepackage{pdflscape}
\usepackage{float}
\usepackage{cancel}
\usepackage{setspace}
\usepackage{graphicx}
\usepackage{array}
\usepackage{hyperref}
\usepackage{appendix}
\usepackage{amssymb}
\usepackage{epstopdf}
 \usepackage{multirow}
\usepackage{epsfig}
\usepackage{subfigure, amsfonts, amsmath}
\usepackage{longtable}
\usepackage{lscape}
\usepackage{subfigure}
\usepackage{amssymb}
\usepackage{yhmath}
\usepackage[english]{babel}
\usepackage[latin1]{inputenc}
\usepackage[T1]{fontenc}
\usepackage{amsthm}
\usepackage{wasysym}
\usepackage{tikz,pgflibraryshapes}
\usepackage{pstricks,pst-plot,pstricks-add}
\usetikzlibrary{arrows}%

\newtheorem{theorem}{Theorem}

\def\XXint#1#2#3{{\setbox0=\hbox{$#1{#2#3}{\int}$}
     \vcenter{\hbox{$#2#3$}}\kern-.5\wd0}}

\renewcommand{\eqref}[1]{Eq.~(\ref{#1})}
\newcommand{\captionfonts}{\footnotesize}

\makeatletter  
\long\def\@makecaption#1#2{%
  \vskip\abovecaptionskip
  \sbox\@tempboxa{{\captionfonts #1: #2}}%
  \ifdim \wd\@tempboxa >\hsize
    {\captionfonts #1: #2\par}
  \else
    \hbox to\hsize{\hfil\box\@tempboxa\hfil}%
  \fi
  \vskip\belowcaptionskip}
\makeatother   

\begin{document}

\title{ON THE THEORY OF SLOWING DOWN GRACEFULLY}
\author{J\"urg Fr\"ohlich and Zhou Gang}

\maketitle
\begin{center}
\section*{Abstract}
\end{center}
\textit{"A moving body will come to rest as soon as the force pushing it no longer acts on it
in the manner necessary for its propulsion."}\begin{flushright}(Aristotle)\end{flushright}
\bigskip
\bigskip

We discuss the transport of a tracer particle through the Bose Einstein condensate of a
Bose gas. The particle interacts with the atoms in the Bose gas through two-body interactions. In
the limiting regime where the particle is very heavy and the Bose gas is very dense, but very weakly
interacting (``mean-field limit"), the dynamics of this system corresponds to classical Hamiltonian
dynamics. We show that, in this limit, the particle is decelerated by emission of gapless modes into
the condensate (Cerenkov radiation). For an ideal gas, the particle eventually comes to rest. In an
interacting Bose gas, the particle is decelerated until its speed equals the propagation speed of the
Goldstone modes of the condensate. This is a model of ``Hamiltonian friction". It is also of interest
in connection with the phenomenon of ``decoherence" in quantum mechanics. It is based on work we have carried out in collaboration with D Egli, IM Sigal and A Soffer.

\newpage
\section{Introduction}

We review some recent results on the effective dynamics of particles moving through a dispersive wave medium, such as the Bose-Einstein condensate of a very dense quantum gas at very low temperature. The motion of the particles exhibits friction as long as their speed is larger than the propagation speed of waves in the medium. Friction is caused by the emission of Cerenkov radiation. This phenomenon provides an example of recent mathematical results in transport theory. A \textit{key problem} in transport theory consists in deriving the effective dynamics of experimentally observable degrees of freedom interacting with unobserved degrees of freedom of a macroscopically large environment from an underlying fundamental (Hamiltonian or unitary quantum) dynamics of the entire system; as indicated in the table below.
\bigskip
\bigskip

\begin{center}
\begin{tabular}{|p{6cm}|p{6cm}|}
\hline
\tikz[remember picture] \node[inner sep=0pt,outer sep=5pt] (n1) {\textbf{Fundamental Dynamics}}; &
      {\tikz[remember picture]\node[inner sep=0pt,outer sep=5pt]  (n2) {\textbf{Effective Dynamics}};} \\
      \hline
Hamiltonian dynamics of point particles and waves - celestial mechanics;& Dissipative dynamics of particles moving through a medium;\\
 Schr\"odinger equation, mechanics of atom gases;... &  friction, diffusion, Brownian motion, (quantum) Boltzmann equation, dynamics of viscous fluids,...\\ \hline
    \end{tabular}
\begin{tikzpicture}[remember picture,overlay]
  \draw[->,red,thin,>=stealth']  (n1.north) to [bend left] node[above]{$?$} (n2.north);
\end{tikzpicture}
\end{center}

The key problem alluded to, above, concerns the passage from the left to the right column in this table. Typically, one might study a system consisting of an observable tracer particle, or a whole tribe of such particles, coupled to unobserved or only crudely observed degrees of freedom of a medium such as the electromagnetic field in an optically dense material, sound waves in a crystal or the Goldstone modes of an atom gas exhibiting Bose-Einstein condensation. At low temperatures, the emission of Cerenkov radiation by the particle(s) into the medium is the basic mechanism causing friction of the particles' motion. When the medium is, initially, in thermal equilibrium at some temperature $T >0$, it exhibits thermal noise that causes a diffusive type of motion of the particle(s) around its (their) mean trajectory.\\
The problem of making these remarks precise by identifying the correct effective dynamics of the particle(s) and analyzing its properties leads one to study a variety of interesting, novel phenomena in the realm of non-linear Hamiltonian evolution equations and quantum dynamics for systems with infinitely many degrees of freedom.

\section{Some simple model systems}
\label{s1}
\subsection{ Dynamics of a particle moving through a medium}
\label{s1.1}
We consider a classical particle moving in physical space $\mathbb{E}^{3}$. The Hamilton function of this system is given by
\begin{equation}
\label{1}
H(X,P)=\frac{P^{2}}{2M} + V(X)
\end{equation}
with $X \in \mathbb{E}^{3}$ the position of the particle, $P \in \mathbb{R}^{3}$ its momentum, $M$ its mass, and $V$ the potential of an external force. An example of a potential is
\begin{equation}
\label{2}
V(X)=-F \cdot X
\end{equation}
where $F \in \mathbb{R}^{3}$ is a constant external force pushing the particle; e.g., the gravitational force close to the surface of the earth.\\
The particle is assumed to interact with a wave medium. We consider the example of a Bose gas exhibiting Bose-Einstein condensation. We would like to treat the entire system within the framework of classical Hamiltonian dynamics. This forces us to pass to a limiting regime of the system, the so-called \textit{mean field limit}: The average density of the Bose gas is given by $\rho_N:=N \rho$, $\rho >0$, the strength of two-body interactions among the atoms of the Bose gas is $\lambda_N:=N^{-1} \lambda$, $\lambda >0$; we consider the limiting regime where $N \rightarrow \infty$, with $\rho$ and $\lambda$ kept fixed. In this limit, the state of the gas can be described by a Ginzburg-Landau order parameter, $\psi(x) \in \mathbb{C}$, where $N \vert \psi(x) \vert^{2}$ is the density of the gas at the point $x \in \mathbb{E}^{3}$. In microscopic units, the mass of a quantum-mechanical tracer particle moving through the Bose gas is given by $NM$ (i.e., it becomes very heavy, as $N \rightarrow \infty$) and the external potential is given by $N V(X)$, with $M$ and $V$ fixed. The limit $N \rightarrow \infty$ corresponds to a "classical limit" in which the quantum dynamics of the  system approaches some classical Hamiltonian dynamics.\\
The Ginzburg-Landau order parameter of the Bose gas, $\psi(x)$, and its complex conjugate, $\bar{\psi}(x)$, are interpreted as complex coordinates of an infinite-dimensional affine phase space, $\Gamma_{BG}:=H^{1}_{\rho}(\mathbb{R}^{3})$, (a certain complex Sobolev space). The standard symplectic structure on $\Gamma_{BG}$ leads to the following \textit{Poisson brackets }:
\begin{equation}\label{3}
 \begin{split}
\lbrace \psi(x), \psi(y)\rbrace&=\lbrace \bar{\psi}(x), \bar{\psi}(y) \rbrace=0\\
\lbrace \psi(x), \bar{\psi}(y)\rbrace&=i \delta(x-y)
\end{split}
 \end{equation}

Quantum mechanically, $\psi$ and $\bar{\psi}$ would be interpreted as annihilation- and creation operators, and the Poisson brackets are replaced by commutators, $\left[\cdot, \cdot \right]$, with $\left[\cdot, \cdot \right] \sim -\frac{i}{N} \lbrace \cdot, \cdot \rbrace$.

In the mean-field limit, $N \rightarrow \infty$, the following Hamilton functional is appropriate to describe a non-relativistic Bose gas:
\begin{eqnarray}
\label{4}
\mathcal{H}_{BG}(\psi, \bar{\psi})=\int d^{3}x \text{ } \left[ \frac{1}{2m} \vert \nabla \psi(x) \vert^{2} + \lambda \int  d^{3}y \text{ } ( \vert  \psi(x) \vert^2 -\rho) \Phi(x-y) ( \vert  \psi(y) \vert^2 -\rho) \right]
\end{eqnarray}
where $\Phi$ is a two-body potential of short range and, for reasons of thermodynamic stability, of \textit{positive type},  $\lambda >0$ is a coupling constant and $\rho>0$ is an average (rescaled) density.  We consider a regime where the $U(1)$-symmetry
\begin{equation}
\psi(x) \mapsto e^{i \theta} \psi(x), \text{ } \bar{\psi}(x) \mapsto e^{-i \theta} \bar{\psi}(x), \text{ } \theta \in \mathbb{R}
\end{equation}
is spontaneously broken. It is convenient to impose symmetry-breaking boundary conditions
\begin{equation}
\label{6}
\overset{(-)}{\psi}(x)=\sqrt{\rho} + \overset{(-)}{\beta}(x)
\end{equation}
with
$$\beta(x) \rightarrow 0, \text{ as } \vert x \vert \rightarrow \infty$$\\
The Poisson brackets between $\beta$ and $\bar{\beta}$ are the same as those in (\ref{3}). In terms of the variables $\beta$ and $\bar{\beta}$, the Hamilton functional is given by
\begin{equation}
\label{7}
\begin{split}
\mathcal{H}_{BG}(\beta, \bar{\beta})=&\int d^{3}x \text{ } \Bigl[ \frac{1}{2m} \vert \nabla \beta(x) \vert^{2} + \\
&\lambda \int  d^{3}y \text{ } \left( \vert  \beta(x) \vert^2 + 2 \sqrt{\rho}  \text{ } \Re \beta(x) \right) \Phi(x-y)  \left( \vert  \beta(y) \vert^2 + 2 \sqrt{\rho} \text{ } \Re \beta(y) \right)\Bigr]
\end{split}
\end{equation}

The interaction between the tracer particle and the Bose gas is given by
\begin{equation}
\label{8}
\begin{split}
\mathcal{H}_{I}(X; \beta,\bar{\beta}):=&g \int d^{3}x \text{ }  W(X-x) \left( \vert \psi(x) \vert^{2} - \rho \right)\\
=&g \int d^{3}x \text{ }  W(X-x) \left( \vert \beta(x) \vert^{2} + 2 \sqrt{\rho} \text{ }\Re  \beta(x)  \right)
\end{split}
\end{equation}
where $X$ is the position of the tracer particle, $W$ is a spherically symmetric two-body potential of short range, and $g >0$ is a coupling constant. The phase space of the coupled system is given by
\begin{equation}
\label{9}
\Gamma_{\text{tot}}:=\mathbb{R}^{6}_{\text{particle}} \times \Gamma_{BG}
\end{equation}
and its Hamilton functional by
\begin{equation}
\label{10}
\mathcal{H}(X,P;  \beta,\bar{\beta} )=H(X,P)+\mathcal{H}_{BG}(\beta,\bar{\beta} )+\mathcal{H}_{I}(X;\beta,\bar{\beta} )
\end{equation}
where $H$ is as in (\ref{1}), $\mathcal{H}_{BG}$ as in (\ref{7}) and $\mathcal{H}_{I}$ as in (\ref{8}).

\subsection{A list of models of decreasing difficulty}
\label{s1.2}
(1) \underline{G-model}:
This is the model specified by (\ref{9}) and (\ref{10}), with $\rho < \infty$, $\lambda >0$, $g>0$.\\\\
(2) \underline{E-model}: We consider the "\textit{Bogoliubov limit}",
\begin{equation} \label{11} \left.\begin{aligned}
\rho&\rightarrow \infty, \text{ }\lambda \rightarrow 0,\text{ } \lambda \rho=\text{const}:=\frac{\kappa}{4} \\
g&\rightarrow 0,\text{ } g \sqrt{\rho}=: \nu, \text{ } \kappa \text{ and } \nu \text{ fixed}\end{aligned} \right\} \end{equation}
In this limit,
\begin{equation*}
\mathcal{H}(X,P; \beta,\bar{\beta})=H(X,P)+\mathcal{H}^{(E)}_{BG}(\beta,\bar{\beta})+\mathcal{H}^{(E)}_{I}(X;\beta,\bar{\beta} )
\end{equation*}
where
\begin{equation}
\label{12}
\begin{split}
\mathcal{H}^{(E)}_{BG}(\beta,\bar{\beta})&=\int d^{3}x \text{ } \Bigl[ \frac{1}{2m} \vert \nabla \beta(x) \vert^{2} +
\kappa \int  d^{3}y \text{ }  \Re  \beta(x)  \text{ }  \Phi(x-y) \text{ } \Re \beta(y) \Bigr]\\
\mathcal{H}^{(E)}_{I}(X; \bar{\beta},\beta)&= 2 \nu \int d^{3} x \text{ }W(X-x) \text{ }  \Re  \beta(x)
\end{split}
\end{equation}
It is useful to introduce real coordinates,
\begin{equation}
\beta(x)=\frac{1}{\sqrt{2}} \left(\varphi(x) +i \pi(x)\right)
\end{equation}
with
\begin{equation*}
\lbrace \varphi , \varphi \rbrace=\lbrace \pi , \pi \rbrace=0, \text{ } \lbrace \varphi(x) , \pi(y) \rbrace=\delta(x-y)
\end{equation*}\\
Then
\begin{equation}
\mathcal{H}^{(E)}_{BG}=\int d^{3}x \text{ } \Bigl[ \frac{1}{4m} \left( \nabla \pi  \right)^{2} (x) + \left( \nabla \varphi  \right)^{2} (x) + \frac{\kappa}{2} \int  d^{3}y \text{ }  \varphi(x) \Phi(x-y) \varphi(y)  \Bigr]
\end{equation}\\
and
\begin{equation}
\mathcal{H}^{(E)}_{I}=\sqrt{ 2} \nu \int d^{3} x \text{ }W(X-x)  \varphi(x)
\end{equation}

The equation of motion for the Bose gas ($g=0$) are
\begin{eqnarray*}
\dot{\varphi}&=&\lbrace\mathcal{H}, \varphi \rbrace=\frac{1}{2m} \Delta \pi\\
\dot{\pi}&=&\lbrace\mathcal{H}, \pi \rbrace=-\frac{1}{2m} \Delta \varphi + \kappa \Phi *\varphi
\end{eqnarray*}
It follows that
\begin{equation}
\ddot{\varphi} =\frac{1}{2m} \Delta \dot{\pi}=-\frac{1}{4m^2}  \Delta^{2} \varphi + \frac{\kappa}{2m} \Delta(\Phi) * \varphi
\end{equation}
which is a wave equation for $\varphi$ with constant coefficients that can be solved by Fourier transformation.\\ One concludes that the frequency, $\Omega$, of a plane wave with wave vector $k \in \mathbb{R}^{3}$ is given by
\begin{equation}
\label{v}
\begin{split}
\Omega(k)&=\vert k\vert \sqrt{\left( \frac{k}{2m} \right)^2+ \frac{\kappa \hat{\Phi}(k)}{2m}}\\
&\approx \vert k\vert \sqrt{ \frac{\kappa \hat{\Phi}(0)}{2m}} \equiv v_{*} \vert k \vert, \text{ } k \approx 0.
\end{split}
\end{equation}
Here, $v_{*}=\sqrt{\frac{\kappa \hat{\Phi}(0)}{2m}}$ is the minimal propagation speed of a Goldstone mode (sound wave) in the Bose-Einstein condensate of the Bose gas.
\bigskip

The equations of motion of the coupled system are given by
\begin{equation}
\label{B}
\begin{split}
M \dot{X} &=P, \text{ } \dot{P}=- \nabla V(X)-\nu \int d^{3} x  \text{ } \nabla W(X-x) \text{ } \Re  \beta(x) \\
i \dot{\beta} (x)&= - \frac{1}{2m} \Delta \beta(x) + \frac{\kappa}{4}\left( \Phi * \Re  \beta  \right)(x) + \nu W^{X}(x)
\end{split}
\end{equation}
where $W^{X}(x):=W(x-X)$.\\\\
(3) \underline{C-model}: $\lambda=0, \text{ } \rho<\infty$, $g>0$; see \cite{EG}.\\\\
(4) \underline{B-model}: $\lambda=0, \text{ } \rho \rightarrow \infty$, with $g\sqrt{\rho}=\nu$ fixed; see \cite{FGS1}, \cite{FGS2}. The equations of motion in the B-model are those given in  (\ref{B}), with $\kappa=0$.

\section{Some special solutions of the equations of motion}
\label{s2}
\subsection{Static solutions: $\dot{X}=\dot{P}=\dot{\beta} \equiv 0$}
\label{s2.1}
Let $X_* \in \mathbb{E}^{3}$ be a critical point of the external potential (with $X_*$ arbitrary if $V \equiv 0$). Let $\beta_*$ be a critical point of the energy functional
\begin{equation*}
\begin{split}
\mathcal{E} (X_*; \beta,\bar{\beta})&:=\int d^{3}x \text{ } \Bigl[ \frac{1}{2m} \vert \nabla \beta(x) \vert^{2} + g W(X_*-x) \left( \vert \beta(x) \vert^{2} + 2 \sqrt{\rho} \text{ }\Re \beta(x) \right) \\
&+ \lambda \int d^{3} y \text{ } \left( \vert \beta(x) \vert^{2} + 2 \sqrt{\rho}  \text{ }\Re  \beta(x) \right) \Phi(x-y) \left( \vert \beta(y) \vert^{2} + 2 \sqrt{\rho} \text{ }\Re  \beta(y) \right)  \Bigr]
\end{split}
\end{equation*}\\
A critical point $\beta_*$  can be constructed with the help of variational calculus. Then $\left(X \equiv X_*, P \equiv 0, \overset{(-)}{\beta}(x) \equiv \overset{(-)}{\beta_*}(x)\right)$ is a static solution of the equations of motion. If $X_*$ is a non-degenerate local minimum of $V$ one expects that there exist solutions of the equations of motion close to this static solution and converging to it,  as time $t$ tends to $\infty$. \\
In the E-model, $\beta_*$ is the real solution of the elliptic equation
\begin{equation}
\label{coucou}
\left( -\frac{\Delta}{2m} + \frac{\kappa}{4} \Phi *\right) \beta=-\nu W^{X_*},
\end{equation}
which can be solved by Fourier transformation. If $\kappa \hat{\Phi}(0)=\kappa \int d^{3}x \text{ } \Phi(x) >0$ then the solution exhibits exponential decay in $x$.

\subsection{Traveling waves for $V \equiv 0$ }
\label{s2.2}
\underline{Ansatz}
\begin{equation} \label{tr} \left.\begin{aligned}
&X_t=vt+X_0, \text{ } P_t \equiv Mv, \text{ }v \in \mathbb{R}^{3}\\
&\beta_t (x)=\gamma_v(x-X_t)\end{aligned} \right\} \end{equation}
Let $W$ be smooth, spherically symmetric and of rapid decay at $\infty$. We consider the E-model, with $v_*=\sqrt{\frac{\kappa \hat{\Phi}(0)}{2m}}>0$; see (\ref{v}). Then, for $\vert v \vert \leq v_c \apprle v_*$ the equations of motion (\ref{B}) have a traveling wave solution of the form (\ref{tr}), and
\begin{center}
$\gamma_v$ decays exponentially fast at $\infty$, for $\vert v \vert <v_c$; $\gamma_v$ has power-law decay at $\infty$, for $\vert v \vert =v_c$.
\end{center}
As $\vert v \vert \rightarrow 0$, $\gamma_v$ converges to $\beta_*$ (see (\ref{coucou})), with $X_*=X_0$. If $\Phi(x)=\delta(x)$ then $v_c=v_*$.\\
The proofs of these claims follow from easy calculations, using Fourier-transformation.\\
For $\vert v \vert >v_c \approx v_*$, the equation $$\frac{k^{2}}{2m}+ \frac{\kappa}{4} \hat{\Phi}(k)-v \cdot k=0$$
has real solutions, $k=k_v \in \mathbb{R}^{3}$, with $k_v \parallel v$. Then there do \textit{not} exist any traveling wave solutions for $V \equiv 0$. If the ansatz (\ref{tr}) is plugged into the equations of motion one finds that $\dot{P} \neq 0$, with $\dot{P}$ anti-parallel to $v$; i.e., a non-zero friction force, $F_v$, arises. This force is caused by emission of Cerenkov radiation by the particle, and its precise form can be inferred from \textit{Fermi's Golden Rules}. \\
If $\kappa =0$ (i.e., $v_*=0$) the expression for $F_v$ is very simple:
\begin{equation}
\label{Fv}
F_v=-\text{const } v \frac{1}{\vert v \vert^{3}} \int_{0}^{(2 \pi v)^{2}} d \rho \text{ } \rho \vert \hat{W}(\sqrt{\rho}) \vert^{2},
\end{equation}
where the constant is proportional to $\nu$. The behaviour of $\vert F_v \vert$ is indicated in the following figure.

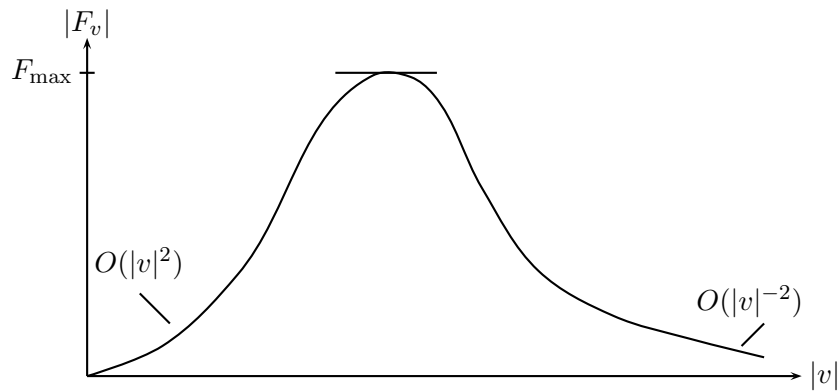
\begin{center}
\newcommand{\imagevecteurx}{\psline{-}(0,0)(1.35,0)}
\newcommand{\imagevecteurxxx}{\psline{-}(0,0)(0.2,0)}
\newcommand{\imagevecteurxx}{\psline{->}(0,0)(9.5,0)}
\newcommand{\imagevecteury}{\psline{->}(0,0)(0,4.5)}
\newcommand{\imagevecteuryy}{\psline{-}(0,0)(0.4,-0.4)}
\newcommand{\imagevecteuryyy}{\psline{-}(0,0)(-0.3,-0.3)}

\begin{pspicture}(0,0)(5,5)
\pscurve(-2,0)(-1,0.4)(0,1.35)(1.8,4)(2.3,3.98)(2.6,3.75)(3.25,2.5)(4,1.4)(5,0.8)(6,0.5)(7,0.25)
\rput(-2,0){\imagevecteury}
\rput(-2,0){\imagevecteurxx}
\uput{0.1}[90](-2,4.4){$\vert F_v \vert$}
\uput{0.1}[90](-2.6,3.8){$ F_{\max}$}
\uput{0.1}[0](7.5,0){$\vert v \vert$}

\uput{0.1}[0](6,1){$O(\vert v \vert^{-2})$}
\uput{0.1}[0](-2,1.5){$O(\vert v \vert^{2})$}

\rput(-2.1,4.035){\imagevecteurxxx}
\rput(1.3,4.035){\imagevecteurx}

\rput(-1.3,1.1){\imagevecteuryy}
\rput(7,0.7){\imagevecteuryyy}

\end{pspicture}
\end{center}

Our discussion suggests to proceed to study \textit{forced} traveling waves, with $V(X)=-F \cdot X$, for a non-vanishing external force $F \in \mathbb{R}^{3}$.
\subsection{Forced traveling waves}
\label{s2.3}
We study the motion of a tracer particle under the influence of  a constant external force $F \in \mathbb{R}^{3}$, with $V(X)=-F \cdot X$. For simplicity, we consider the B- or the E-model.
\begin{theorem} \cite{EFGS}
Suppose that $W$ is smooth, spherically symmetric and of rapid decay at infinity. Then there exists a positive constant $F_{\text{max}} < \infty$ such that
\bigskip

(1) if $\vert F \vert <F_{\text{max}}$ the equations of motion (\ref{B}) of the B- or E-model have \textit{two} traveling wave solutions of the form (\ref{tr}) with velocity $v \parallel F$, $\vert v \vert >v_c \approx v_*$;
\bigskip

(2) if $\vert F \vert = F_{\text{max}}  $ there exists a unique traveling wave solution with $v \parallel F$, $\vert v \vert >v_c$;
\bigskip

(3) if $\vert F \vert >F_{\text{max}}$ there do not exist any traveling wave solutions of the form (\ref{tr}).
\end{theorem}
\underline{Remarks}. To prove this theorem we must look for solutions of the equations of motion of the form $X_t=vt+X_0$, $P_t \equiv Mv$, $\beta_t(x)=\gamma_v (x- X_t)$.\\
The equations of motion (\ref{B}) then imply that
\begin{equation}
\label{mo}
\begin{split}
&- \nabla V(X) \equiv F =\nu \int d^{3} x \text{ }(\nabla W)(x) \text{ }\Re  \gamma_v(x) \\
&-i v \cdot  \nabla \gamma_v= -\Delta \gamma_v + \frac{\kappa}{4} \left( \Phi * \Re  \gamma_v \right)(x)+ \nu W(x)
\end{split}
\end{equation}\\
If $\vert v \vert >v_c$ the Fourier transform, $\hat{\gamma}_v$, of the second equation in (\ref{mo}) is singular on a spherical surface in k-space, and it follows that the first equation in (\ref{mo}) does not have any solutions unless $\vert F \vert>0$, as discussed in section \ref{s2.2}. \\
The right side of the first equation in (\ref{mo}) is a function of $\vert v \vert$ that vanishes, for $\vert v \vert <v_c$, tends to $0$, as $\vert v \vert \rightarrow \infty$, and has a unique maximum at some value, $v^{*}$, of $\vert v \vert$. It is quadratic in $W$.

From these observations the proof of the theorem stated above can easily be inferred. For $\vert v \vert >v_c$, $\vert F \vert \neq 0$, the solution, $\gamma_v$, of the second equation describes a conical wave whose tip is at the position, $X_t$, of the particle.

\subsection{Significance of special solutions}
\label{s2.4}
The construction of special solutions, as outlined in Sects \ref{s2.1}, \ref{s2.2}  and \ref{s2.3}  is elementary, and one may wonder why such solutions are of any interest, at all. The reason is that one expects that rather general classes of (time-dependent, non-stationary) solutions of the equations of motion approach the special solutions constructed above, as time $t$ tends to $\infty$, (in a mathematically precise sense). We briefly discuss this expectation in several situations.

(1) Let us suppose that $V(X)$ has a non-degenerate (quadratic) minimum, $V(X_*) <0$, at a point $X_* \in \mathbb{E}^{3}$, with $V(X) \rightarrow 0$, as $\vert X \vert \rightarrow \infty$. We consider the G-model and choose initial conditions, $(X_0, P_0, \beta_0)$, whose total energy, $\mathcal{H}(X_0,P_0; \beta_0,\bar{\beta}_0)$, is strictly \textit{negative}. Then we expect that the solution, $(X_t, P_t,\beta_t)$, of the Hamiltonian equations of motion corresponding to the Hamilton functional $\mathcal{H} (P,X;\beta,\bar{\beta})$ in (\ref{10}) approaches a static solution, $X_t \rightarrow X_*$, $P_t \rightarrow 0$, $\beta_t \rightarrow \beta_*$, as $t \rightarrow \infty$, where $\beta_*$ is a minimizer of the energy functional $\mathcal{E}(X_*; \beta,\bar{\beta})$ introduced in section \ref{s2.1}, provided the coupling constant $g$ is sufficiently small. (If the potential $V$ is as specified above, but $V(X) \rightarrow \infty$, as $\vert X \vert \rightarrow \infty$, then one expects that the static solution $(X_*, P=0,\beta_*)$ is asymptotically stable without further assumptions on initial conditions of finite total energy and on the coupling constant $g$.)

(2) We consider the B- or E-model, with $V(X) \equiv 0$. We choose initial conditions, ($X_0,P_0,\beta_0$), of energy $\mathcal{H}(X_0,P_0;  \beta_0,\bar{\beta}_0)$ (see (\ref{12})) not "much larger" than $\frac{M}{2} v^{2}_c$ (with $v_c=v_*=0$, for the B-model) and $g$ small enough. Then we expect that the solution, $(X_t,P_t,\beta_t)$, of the equations of motion (\ref{B}), with the given initial conditions, approaches a traveling wave solution (\ref{tr}), with $\vert v \vert \leq v_c$, as $t \rightarrow \infty$; (convergence of $\beta_t$ is understood locally, in arbitrary balls of finite radius around the position, $X_t$, of the tracer particle).

(3) We consider the B- or E-model, but with $V(X)=- F \cdot X$, $0<\vert F \vert < F_{\text{max}}$. Let $v_{<}>0$ and $v_>>v_<$ be the two speeds for which the equations in (\ref{mo}) have solutions, with $\vert v \vert =v_<$ or $\vert v \vert =v_>$. We first discuss the stability of solutions with $\vert v \vert =v_<$. We expect that a solution of the equations of motion (\ref{B}) corresponding to an initial condition $(X_0,P_0,\beta_0)$, with $X_0 \in \mathbb{E}^{3}$ arbitrary, $\vert P_0 \vert <<Mv_<$, $\mathcal{H}^{(E)}_{BG}( \beta_0,\bar{\beta}_0)$ small enough, converges to a traveling wave solution of the type discussed in the theorem of section \ref{s2.3}, with $\vert v \vert =v_<$, as time $t \rightarrow \infty$, provided $g$ is small enough.

In contrast, if $\vert P_0 \vert \apprge Mv_>$,  $\mathcal{H}^{(E)}_{BG}( \beta_0,\bar{\beta}_0)$ small, the solution $(X_t, P_t,\beta_t)$ of the equations of motion (\ref{B}) will in general not approach a traveling wave solution with $\vert v \vert =v_>$. Rather, we expect that $\vert P_t \vert $ diverges, as $t \rightarrow \infty$. (Similar behaviour is expected if $\vert F \vert >F_{\text{max}}$.)\\ Most of these expectations remain unproven, except for expectation (2) in the context of the B-model.

\section{A theorem deserving this name}
\label{s3}
In this section, we consider the B- and C-models, which are the only models for which non-trivial results have been established, so far; see \cite{FGS1}, \cite{FGS2},\cite{EG}. We state our main result for the B-model; but a very similar result has also been proven for the C-model. We propose to study solutions of the equations of motion (\ref{B}), for $\kappa=0$.

We choose units such that $M=m=1$. The two-body potential $W$ is assumed to be smooth, spherically symmetric and of rapid decay at $\infty$, with $\vert \hat{W}(0) \vert=(2 \pi)^{-3/2} \vert \int d^{3} x \text{ } W(x) \vert=1$. The wave field $\beta$ is normalized appropriately. The external potential $V$ is assumed to vanish identically. Then the equations of motion (\ref{B}) take the form
\begin{equation}
\label{mo3}
\begin{split}
 \dot{X}_t &=P_t, \text{ } \dot{P}_t=\nu \int d^{3} x  \text{ } (\nabla W^{X_t})(x) \text{ }\Re  \beta_t (x) \\
i \dot{\beta} _t (x)&= - \frac{1}{2} (\Delta \beta_t)(x) +  W^{X_t}(x)
\end{split}
\end{equation}
where $ W^{X_t}(x):=W(x-X_t)$, $\nu=O(1)$ a constant. Note that the equations (\ref{mo3}) have static solutions
$$X_t \equiv X_*, \text{ } P_t \equiv 0, \text{ }\beta_t(x) \equiv \beta_*(x)=2(\Delta^{-1} W^{X_*})(x)$$
indexed by $X_*$, and we propose to analyze their asymptotic stability. Our main result is the following theorem.

\begin{theorem} \cite{FGS2}
Under the assumptions described above, there exists an interval $I \supseteq \left[0,0.66 \right) \supset \left[ 0, \frac{1}{2} \right]$ such that, for any $\delta \in I$, there is some $\epsilon_0 = \epsilon_{0}(\delta) >0$ with the property that if
$$\vert P_0 \vert, \text{ } \vert \vert \langle x \rangle^{3} \beta_0 \vert \vert_2, \text{ } \vert \vert \nabla \beta_0 \vert \vert_2  <\epsilon_0 $$
where $\langle x \rangle:=\sqrt{1+ \vert x \vert^{2}}$, then
\begin{equation}
\vert P_t \vert \leq c \langle t \rangle^{-\frac{1}{2}- \delta}, \text{ as } t \rightarrow \infty
\end{equation}
for some finite constant $c$, and
$$\lim_{t \rightarrow \infty} \vert \vert \beta_t-2 \Delta^{-1} W^{X_t} \vert \vert_{\infty}=0$$
\end{theorem}
\underline{Remarks}:\\
(1) If $\delta$ is chosen to  be larger than $\frac{1}{2}$ (note that I contains such values of $\delta$) then
$$X_t =X_0+ \int_{0}^{t} P_s\text{ }  ds \rightarrow X_*, \text{ as } t \rightarrow \infty$$
with $\vert X_* \vert < \infty$.\\\\
(2) We suspect that $\vert P_t \vert \sim t^{-\frac{1}{2}- \delta}$, as $t \rightarrow \infty$, for an exponent $\delta > \frac{1}{2}$ \textit{independent} of initial conditions, as long as $\vert P_0 \vert$, $\vert \vert \langle x \rangle^{3}  \beta_0 \vert \vert_2$, $\vert \vert \nabla \beta_0 \vert \vert_2$, are finite. However, this has not been proven, so far. (If the condition that $\vert \vert \langle x \rangle^{3}  \beta_0 \vert \vert_2 < \infty$ is dropped then $\delta$ may depend on initial conditions.\\
For, consider the equation
\begin{equation}
\label{fvv}
\dot{v}_t=F_{v_t}
\end{equation}
with $F_v$ as in (\ref{Fv}). This corresponds to $\beta_t(x)=\gamma_{v_t} (x-X_t)$. The solution of (\ref{fvv}) behaves like $\vert v_t \vert \sim \langle t \rangle^{-1}$, i.e., $\delta=\frac{1}{2}$. However, $\vert \vert \beta_t \vert \vert_2$ is divergent!)\\\\
(3) The theorem stated above has interesting applications in the analysis of \textit{decoherence} in systems of a heavy quantum-mechanical particle moving through a Bose-Einstein condensate.

Although the initial moves in the proof of the theorem described above are quite natural and fairly standard, the technical details are suprisingly tedious and involve exhibiting very subtle cancellations among terms that do not, a priori, exhibit adequate decay, as time $t \rightarrow \infty$, in order to "close the estimates".

It would be of considerable interest to study systems of finitely many tracer particles moving through a Bose-Einstein condensate, or a gas of such particles. It is not very hard to come up with plenty of interesting and plausible looking conjectures concerning the behaviour of such systems. But it appears to be very difficult to rigorously prove these conjectures.\\\\
\small
\underline{Note}:
For further results on friction, see also \cite{bruneau, caprino, KM}.

\normalsize
\begin{acknowledgements}
We thank D.Egli, I.M. Sigal and A. Soffer for many enlightening discussions.  We are indebted to B.Schubnel for carefully reading the manuscript and much help with the typesetting.
\end{acknowledgements}

\bibliographystyle{plain}


\begin{thebibliography}{1}

\bibitem{bruneau}
L.~Bruneau and S.~De~Bi{\`e}vre.
\newblock A hamiltonian model for linear friction in a homogeneous medium.
\newblock {\em Communications in mathematical physics}, 229(3):511--542, 2002.

\bibitem{caprino}
S.~Caprino, C.~Marchioro, and M.~Pulvirenti.
\newblock Approach to equilibrium in a microscopic model of friction.
\newblock {\em Communications in mathematical physics}, 264(1):167--189, 2006.

\bibitem{EFGS}
D.~Egli, Fr\"ohlich J., Zhou G., and Sigal I.M.
\newblock Papers in preparation.

\bibitem{EG}
D.~Egli and G.~Zhou.
\newblock Some hamiltonian models of friction ii.
\newblock {\em Arxiv preprint arXiv:1108.5545}, 2011.

\bibitem{FGS2}
J.~Fr\"ohlich, Z.~Gang, and A.~Soffer.
\newblock Friction in a model of hamiltonian dynamics.
\newblock {\em Arxiv preprint arXiv:1110.6550}, 2011.

\bibitem{FGS1}
J.~Fr{\"o}hlich, Z.~Gang, and A.~Soffer.
\newblock Some hamiltonian models of friction.
\newblock {\em Journal of Mathematical Physics}, 52:083508, 2011.

\bibitem{KM}
DL~Kovrizhin and LA~Maksimov.
\newblock Cherenkov radiation of a sound in a bose condensed gas.
\newblock {\em Physics Letters A}, 282(6):421--427, 2001.

\end{thebibliography}

\end{document}